\documentclass[12p]{article}
\usepackage{latexsym}
\usepackage{color}
\usepackage{amsmath}
\usepackage{epsfig}
\usepackage{hyperref}
 \usepackage{dcolumn}
\newcommand{\ortala}[1]{\begin{center}#1\end{center}}

\newcommand{\sandd}[1]{\left\langle #1\right\rangle}
\newcommand{\sanddr}[1]{\left\langle\left\langle #1\right\rangle\right\rangle_r}
\newcommand{\integ}[3]{{{\underset{#1 }{\overset{#2}{\displaystyle\int}}}#3}}

\newcommand{\summ}[3]{{{\underset{#1 }{\overset{#2}{\displaystyle\sum}}}#3}}

\newcommand{\re}[1]{(\ref{#1})}

\newcommand{\eq}[2]{\begin{equation}\label{#1}  #2\end{equation}}

\newcommand{\paran}[1]{\left(#1\right)}

\newcommand{\sch}[1]{Schrodinger}
\newcommand{\ktur}[2]{\frac{\partial #1}{\partial #2}}

\newcommand{\komb}[2]{\paran{\begin{array}{c} #1 \\ #2 \end{array}}}
\newcommand{\sanddrtek}[1]{\left\langle\left\langle 
#1\right\rangle\right\rangle_{r}}

\usepackage{amssymb}
\usepackage{latexsym}
\begin{document}

\ortala{\large\textbf{Magnetocaloric properties of the binary Ising model with arbitrary spin}}

\ortala{\textbf{\"Umit Ak\i nc\i \footnote{\textbf{umit.akinci@deu.edu.tr}}}}

\ortala{\textit{Department of Physics, Dokuz Eyl\"ul University,
TR-35160 Izmir, Turkey}}

\section{Abstract}

The magnetocaloric  properties of the Ising  binary alloys composed of arbitrary spin values, 
have been determined by using effective field theory. For determining the efficiency of the magnetocaloric effect
in binary alloy, the quantities of interest such as isothermal magnetic entropy change and refrigerant capacity 
have been calculated for various values of spin and concentration. It is shown that, by changing the 
concentration we can tune the magnetocaloric performance. Also,  it has been shown that 
greater refrigerant capacity can be obtained 
for intermediate concentration values, in   comparison with the limiting concentration values.

\section{Introduction}\label{introduction}

Magnetic cooling is one of the broad application areas of the magnetism. 
The magnetocaloric effect (MCE) \cite{ref1,ref2,ref3,ref4} allow us to use the magnetism  in 
cooling or heating applications. 
If the magnetic field on the magnetic material  is changed in an adiabatic way, the material 
can change its temperature. This is due to the balancing between the lattice (Debye term) and magnetic parts 
of the total entropy. Rising magnetic contribution in the total entropy change will yield decreasing the lattice 
contribution.
This means that decreasing lattice vibrations and thus decreasing temperature. Due to the zero total entropy change 
(which comes from the adiabaticity of the process), by controlling magnetic part  of the total entropy, one can 
achieve cooling or heating of the material.

There are many candidate materials of the MCE \cite{ref5}.  
There are some parameters that determine which candidate is more efficient. Some of the parameters are 
isothermal magnetic entropy change (IMEC) $\Delta S_M$,  adiabatic
temperature change $\Delta T_{ad}$ ,and  refrigerant capacity ($RC$) $q$, which are defined below. We can 
say that $Gd$ is still the best candidate material  material for the MCE \cite{ref6}.

Naturally, the search for better materials still continues. Some recent developments can be found in review papers 
\cite{ref7,ref8}. Unfortunately, there are relatively small number of theoretical works devoted to the exploration of 
MCE for given magnetic materials. Diluted ferromagnetic systems \cite{ref9},  magnetic multilayers \cite{ref10}
quantum spin chains \cite{ref11} are some of these studies.

Alloys may be good candidate materials because of the fact that, tunability of the concentration 
may yield obtaining material that has desired MCE properties. Some theoretical efforts 
have been made in this direction. For instance $Gd_{1-x} C_x$ binary alloy \cite{ref12}, 
$Gd_5(Si_xGe_{1-x})_4$ compound \cite{ref13} , $Gd-R$ alloys \cite{ref14}. Investigating 
alloys from a broad framework can produce important results that can be used in experiments or material 
science applications. Thus, the aim of this work is to determine the MCE properties of the binary 
alloys that can be modeled by the Ising model. The effect of the spin value 
(that constitute the alloy) and the
concentration on the MCE properties are calculated. 
For this aim, the paper is organized as follows: In Sec. 
\ref{formulation} we briefly present the model and  formulation. The results and
discussions are presented in Sec. \ref{results}, and finally Sec.
\ref{conclusion} contains our conclusions.

\section{Model and Formulation}\label{formulation}

The composition of the binary alloy system can be represented as 
$A_cB_{1-c}$. Here, the type $A$ atoms have spin value of $s_A$ and the concentration is $c$,
the type $B$ atoms have  concentration $1-c$ and spin value of $s_B$. The distribution of 
the atoms are random. The Hamiltonian of the Ising model for this system can be given as  

 \eq{denk1}{
\mathcal{H}=-J\summ{<i,j>}{}{}\summ{X,Y=A,B}{}{}\xi_i^X \xi_j^Y  S_i^{X} S_j^{Y}
 -\summ{i}{}{}\summ{X=A,B}{}{}\xi_i^X\left[D
\paran{S_i^{X}}^2+H S_i^{X}\right]}
where $S_i^{(X)}, X=A,B$ are the $z$ components of the spin-$s_X$  
operators. The eigenvalues of these operators are 
$s_i^{X}=-s_X, -s_X+1,\ldots, s_X-1, s_X $.
$J>0$ is the   ferromagnetic
exchange interaction between the nearest neighbor spins, $D$ is
the crystal field (single ion anisotropy) and $H$ is the longitudinal magnetic field.  
$\xi_i^X$ 
is the site occupation number of a site $i$, which can take values $0,1$ and 
$\xi_i^A+\xi_i^B=1$ holds. In Eq. \re{denk1}, $<i,j>$ denotes the nearest neighbor pairs 
in the lattice.

Typical effective field approach starts by writing Hamiltonian given in Eq. \re{denk1} as 
single atom part ($H_0^X$) and 
the part that contains rest of the lattice ($H^\prime$). Since the Hamiltonian contains only  $z$ components of 
the spin operators, $H_0^X$ and $H^\prime$ commutes each other. Then 
 the exact identities \cite{ref15} can be used for calculation of the thermodynamic quantities 
$\paran{s_0^{X}}^n$

\eq{denk2}{
\frac{\sanddr{\xi_0^X \paran{S_0^X}^n}}{\sandd{\xi_0^X}_r}=\frac{1}{\sandd{\xi_0^X}_r}
\sanddr{\frac{Tr_0\xi_0^X \paran{S_0^X}^n \exp{\paran{-\beta
\mathcal{H}_0^{X}}}}{Tr_0\exp{\paran{-\beta \mathcal{H}_0^{X}}}}},
}
where $n=1$ represents the magnetic dipol moment per site, 
$n=2$ represents the magnetic quadrupol moment per site and so on.  
Here, $<>$ represents the thermal average, $<>_r$ stands for the configurational average,
$Tr_0$ is the partial trace over the site
$0$, $\beta=1/\paran{k_B T}$, $k_B$ is Boltzmann constant
and $T$ is the temperature. Note that Eq.  \re{denk2} contains number of $\paran{2s_X+1}$ equations for 
sublattice $X$. This means that total number of equations is $\paran{2s_A+2s_B+2}$ for binary alloy.

In Eq. \re{denk2},  $\mathcal{H}_0^{X}$ is the  local field acting on site $0$ if 
the site has $X=A,B$ type atoms. Their expressions are as follows: 

\eq{denk3}{
\mathcal{H}_0^X=
-\xi_0^X s_0^X\left[E_0^{X}+H\right]-\xi_0^X\paran{s_0^X}^2 D,}
where $E_0^{X}$ represents all spin-spin interactions of atom which is the type of $X=A,B$
and it is given by
\eq{denk4}{
E_0^{X}=J\summ{j=1}{z}{}\summ{X=A,B}{}{}\xi_j^X S_j^X.
}

All $Tr_0$ operations in Eq. \re{denk2} can be done easily, since all of the matrices in that equation  
are diagonal. After this operation we can get the expression in a closed form as 
\eq{denk5}{
\frac{\sanddr{\xi_0^X \paran{S_0^X}^n}}{\sandd{\xi_0^X}_r}=
\sanddr{f_n^X\paran{ E_0^{X}}},} where the functions for $n=1,2$ are given by \cite{ref16}

\eq{denk6}{f_1^X\paran{x,H,D}=\frac{\summ{k=-S_X}{S_X}{}k\exp{\paran{\beta D 
k^2}\sinh{\left[\beta k\paran{x+H}\right]}}}{\summ{k=-S_X}{S_X}{}\exp{\paran{\beta D 
k^2}\cosh{\left[\beta k\paran{x+H}\right]}}},
}

\eq{denk7}{f_2^X\paran{x,H,D}=\frac{\summ{k=-S_X}{S_X}{}k^2\exp{\paran{\beta D 
k^2}\cosh{\left[\beta k\paran{x+H}\right]}}}{\summ{k=-S_X}{S_X}{}\exp{\paran{\beta D 
k^2}\cosh{\left[\beta k\paran{x+H}\right]}}}.
} Note that, as explained below, within the approximation schema used in this 
work there is no need to obtain functions  for $n>2$.

By using differential operator technique \cite{ref17}, Eq. \re{denk5} can be 
written as

\eq{denk8}{
\frac{\sanddr{\xi_0^X \paran{S_0^X}^n}}{\sandd{\xi_0^X}_r}=\sanddrtek{e^{E_0^{X}\nabla}}f_n^X(x)|_{x=0},
} where  $\nabla$ represents the differential with respect to $x$. 
The
effect of the differential operator $\nabla$ on an arbitrary function $F$ is
given by
\eq{denk9}{\exp{\paran{a\nabla}}F\paran{x}=F\paran{x+a},} with arbitrary 
constant $a$. By using $E_0^X$ from Eq. \re{denk4} in Eq. \re{denk8} we can expand 
the exponential differential operator by using approximated van der 
Waerden identities \cite{ref18}.  Approximated van der Waerden identities have been
proposed for higher spin problems and given by
\eq{denk10}{
\exp{\paran{a\sigma}}=\cosh{\paran{a\eta}}+\frac{\sigma}{\eta} \sinh{\paran{a\eta}},
} where $\eta^2=\sandd{\sigma^2}$ and $\sigma$ is the spin eigenvalue.  This 
approximation equates $\sigma^{2n}$ to $\sandd{\sigma^2}^n$ and $\sigma^{2n+1}$ to $\sigma\sandd{\sigma^2}^n$.
Since we are dealing only with the quantities $\sigma$ and $\sigma^2$, then
the number of equations in Eq. \re{denk5} reduces to four.  By using Eq. \re{denk10} with the identity 
\eq{denk11}{
e^{\delta x}=\delta e^{x}+1-\delta, \quad 
 \delta=0,1} in Eq. \re{denk8},
we can obtain magnetization ($m_A,m_B$) and quadrupolar moment ($q_A,q_B$) equations as

\eq{denk12}{
m_X=\summ{p=0}{z}{}\summ{q=0}{z-p}{}\summ{r=0}{p}{}\summ{s=0}{z-q-r}{}\summ{t=0}{q+r}{}
C_{pqrst}(-1)^tc^{z-p}\paran{1-c}^p \paran{\frac{m_A}{\eta_A}}^q\paran{\frac{m_B}{\eta_B}}^r 
f_1^x\paran{\left[z-2s-2t\right] J,} 
}
\eq{denk13}{
q_X=\summ{p=0}{z}{}\summ{q=0}{z-p}{}\summ{r=0}{p}{}\summ{s=0}{z-q-r}{}\summ{t=0}{q+r}{}
C_{pqrst}(-1)^tc^{z-p}\paran{1-c}^p \paran{\frac{m_A}{\eta_A}}^q\paran{\frac{m_B}{\eta_B}}^r 
f_2^x\paran{\left[z-2s-2t\right] J,} 
} where $\eta_X^2=q_X$, $X=A,B$ and 
\eq{denk14}{
C_{pqrst}=\frac{1}{2^z}\komb{z}{p}\komb{z-p}{q}\komb{p}{r}\komb{z-q-r}{s}\komb{q+r}{t}.
}

By solving the system of nonlinear equations given by  Eqs.
\re{denk12} and \re{denk13}  with the coefficients given in 
\re{denk14} and the definitions of functions given in Eqs. \re{denk6} and \re{denk7}, we
can obtain the total magnetization ($m$) and quadrupolar moment ($q$) of the system as
\eq{denk15}{
m=cm_A+\paran{1-c}m_B, \quad q=cq_A+\paran{1-c}q_B.
}

The magnetocaloric performance of the magnetic system can be quantified by some quantities. The first one is 
IMEC when maximum applied  longitudinal field is $h_{max}$. This quantity is defined by
\eq{denk16}{
\Delta S_M=\integ{0}{h_{max}}{}\paran{\ktur{m}{T}}_h dh.
} 

The other quantity is RC which is defined by
\eq{denk17}{
q=-\integ{T_1}{T_2}{} \Delta S_M\paran{T}_H dT. 
} 
Note that, RC is amount of heat
that can be transferred from the cold end (at temperature $T_1$)  to
the hot end (at temperature $T_2$ ) in one thermodynamical cycle. This measures
the  suitability of a magnetic material for magnetocaloric purposes. 
Although the chosen of these temperatures are arbitrary, in general 
$T_1$ and $T_2$ are chosen as those temperatures at which
the magnetic entropy change gains the half of its peak value. This temperature range is 
defined as full width at half max (FWHM). This is  
also important quantity of the MCE. This three quantities will be used in this work as a  magnetocaloric
properties of the binary alloy.

\section{Results and Discussion}\label{results}

The following dimensionless parameters are used in this work:

\eq{denk18}{
\quad d=\frac{D}{J}, \quad h=\frac{H}{J}, \quad t=\frac{k_BT}{J}.  
}
For a systematic investigation, we always use $S_A<S_B$. This means that, rising $c$ always corresponds the 
rising concentration of the atoms that have lower spin value. 

In our earlier work \cite{ref19} we have concluded about the magnetocaloric properties of the 
general spin-S Ising model and they were as follows: 

\begin{itemize}
\item For a chosen external magnetic field range, the 
maximum value of the IMEC decreases when the spin value increases, 

 \item For a chosen external magnetic field range, the FWHM and RC
increase when the spin magnitude of the system is increased,  


\end{itemize}

Decreasing maximum value of the IMEC and increasing FWHM and RC by 
rising spin value should manifest itself in binary alloys, when $c$ decreases.  
For the validation of these conclusions,
we depict some typical IMEC with the temperature. This behaviors can be seen
in Fig. \re{sek1} for binary alloy consists of $(S_A,S_B)$, 
(a)$(1,3/2)$ and (b)$(1,3)$ by using $h_{max}=1.0$ in Eq. \re{denk16}.

\begin{figure}[h]\begin{center}
\epsfig{file=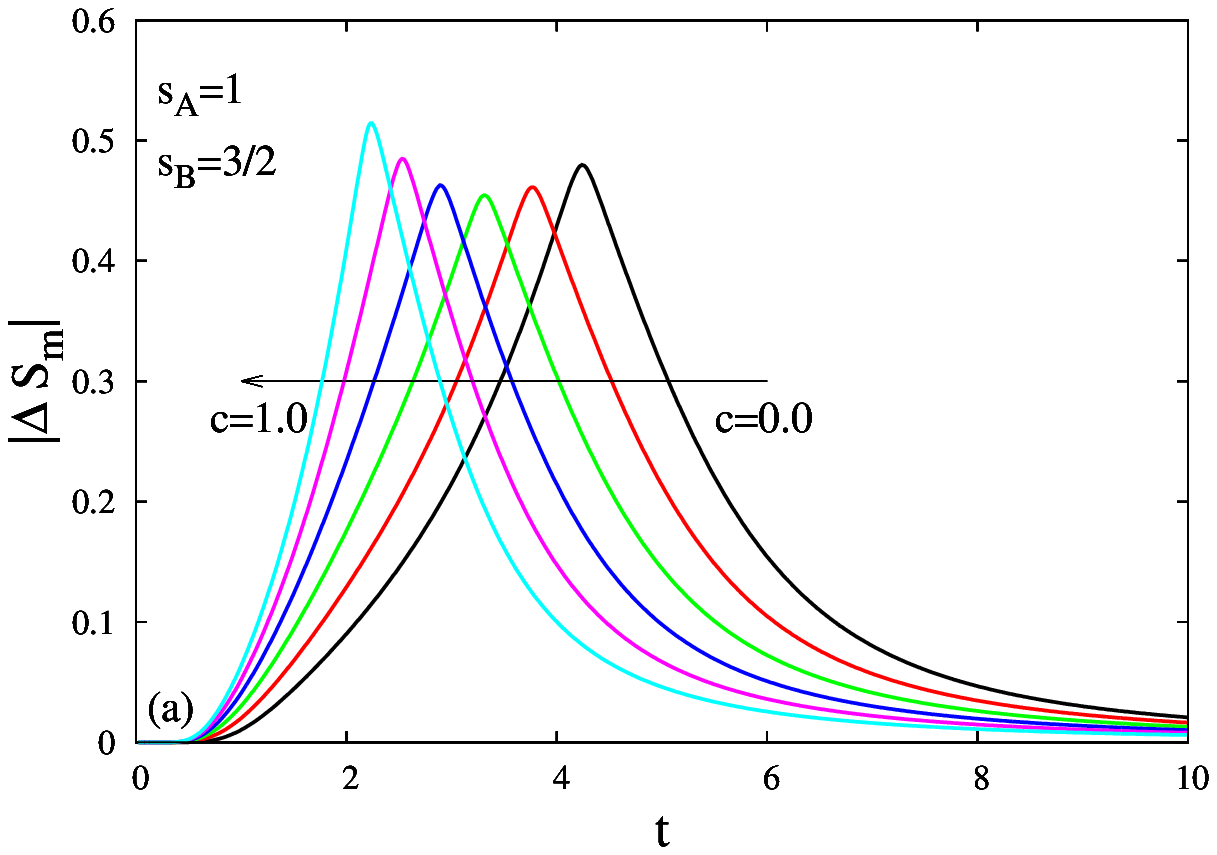, width=6.0cm}
\epsfig{file=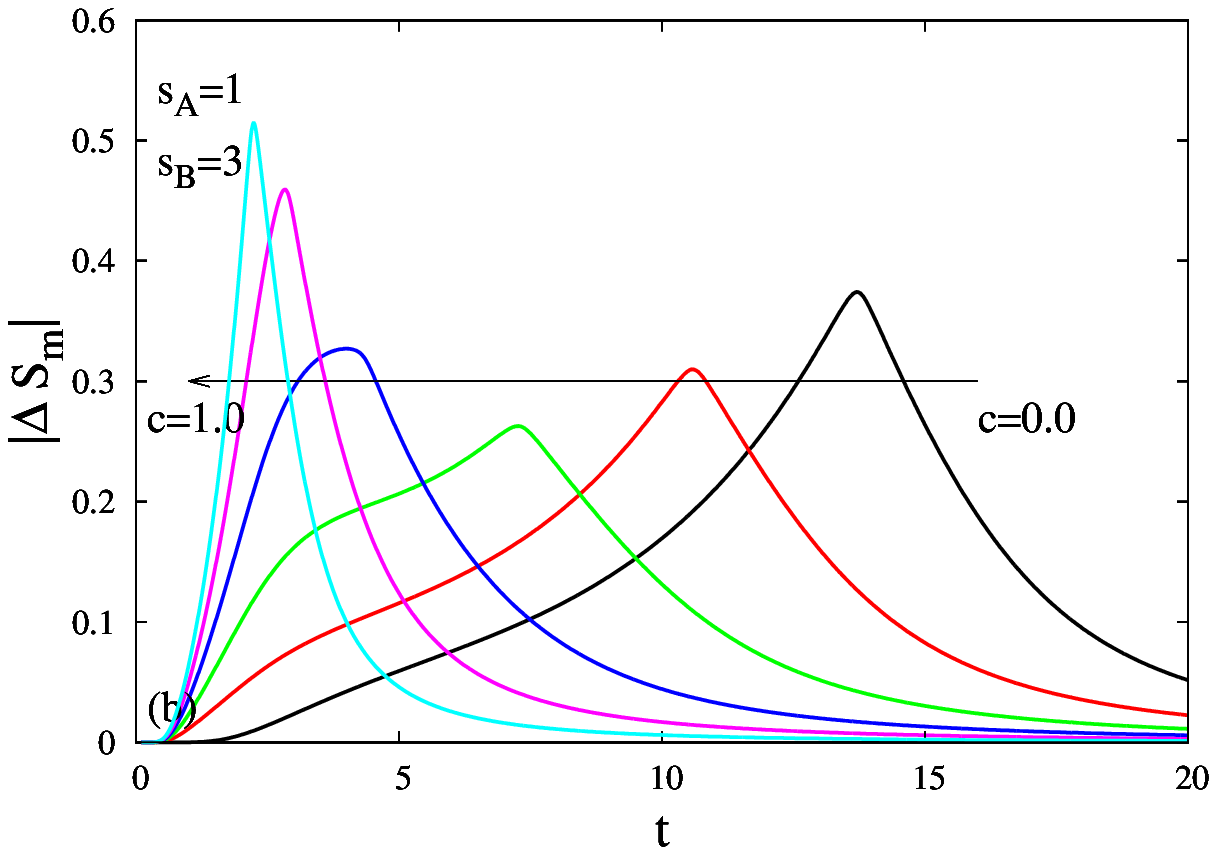, width=6.0cm}
\end{center}
\caption{The variation of the IMEC with the temperature for binary alloy 
(a) $(S_A,S_B)=(1,3/2)$ and (b)$(S_A,S_B)=(1,3)$. For obtaining these curves, $h_{max}=1.0$ has been used 
in Eq. \re{denk16}, and numerical calculations have been performed both for integration and differentiation. 
These curves are depicted for concentration values of $c=0.0,0.2,0.4,0.6,0.8,1.0$.} \label{sek1}
\end{figure}

From Fig. \re{sek1}, general conclusions mentioned above can be seen. But surprisingly,
rising concentration ($c$) of $A$ atoms  
results no monotonic behavior in FWHM and the maximum value of the IMEC. For instance 
the maximum value of the IMEC is decreasing for a while, then increasing by rising  concentration
value which is starting from $c=0$.  This behavior is more evident for larger  values of $S_B$
(compare Fig. \re{sek1} (b) by (a)). On the other hand, $(S_A,S_B)=(1,3)$ binary alloy displays 
broadening behavior in the variation of IMEC with the temperature, for the intermediate concentration values (see for 
curves related to the $c=0.2,0.4,0.6$ in Fig. \re{sek1} (b) ). 
We can inspect these two facts by investigation of the maximum value of the IMEC
and FWHM with concentration
by constructing different binary alloys, i.e. choosing different $(S_A,S_B)$ pairs.

\begin{figure}[h]\begin{center}
\epsfig{file=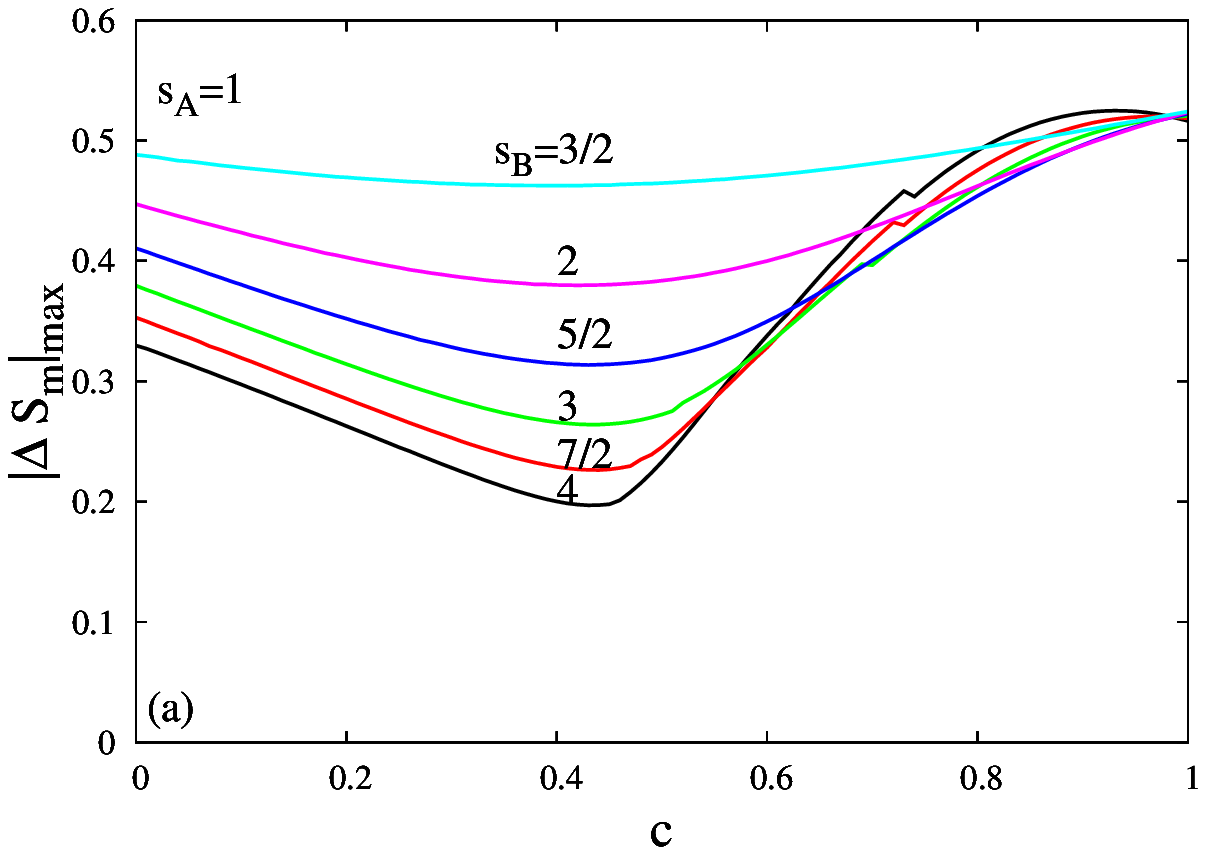, width=6.0cm}
\epsfig{file=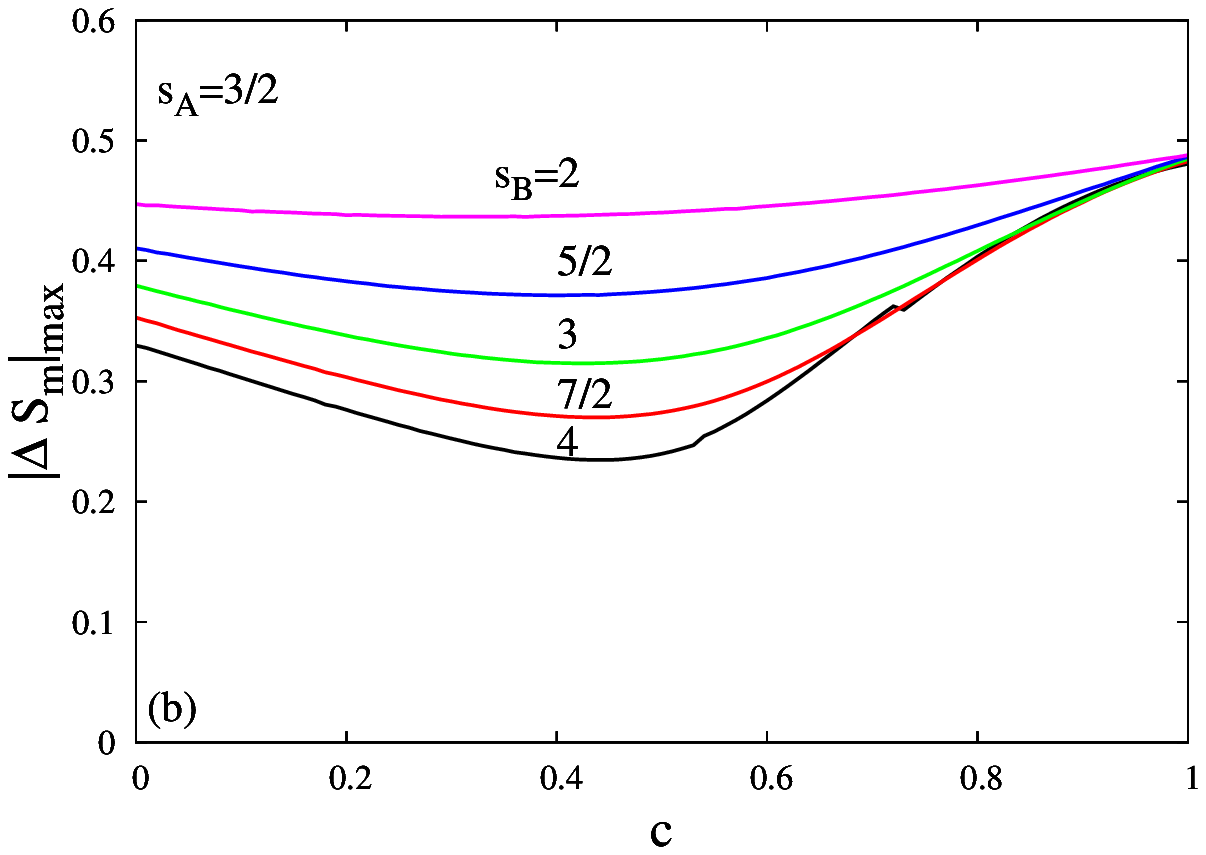, width=6.0cm}
\epsfig{file=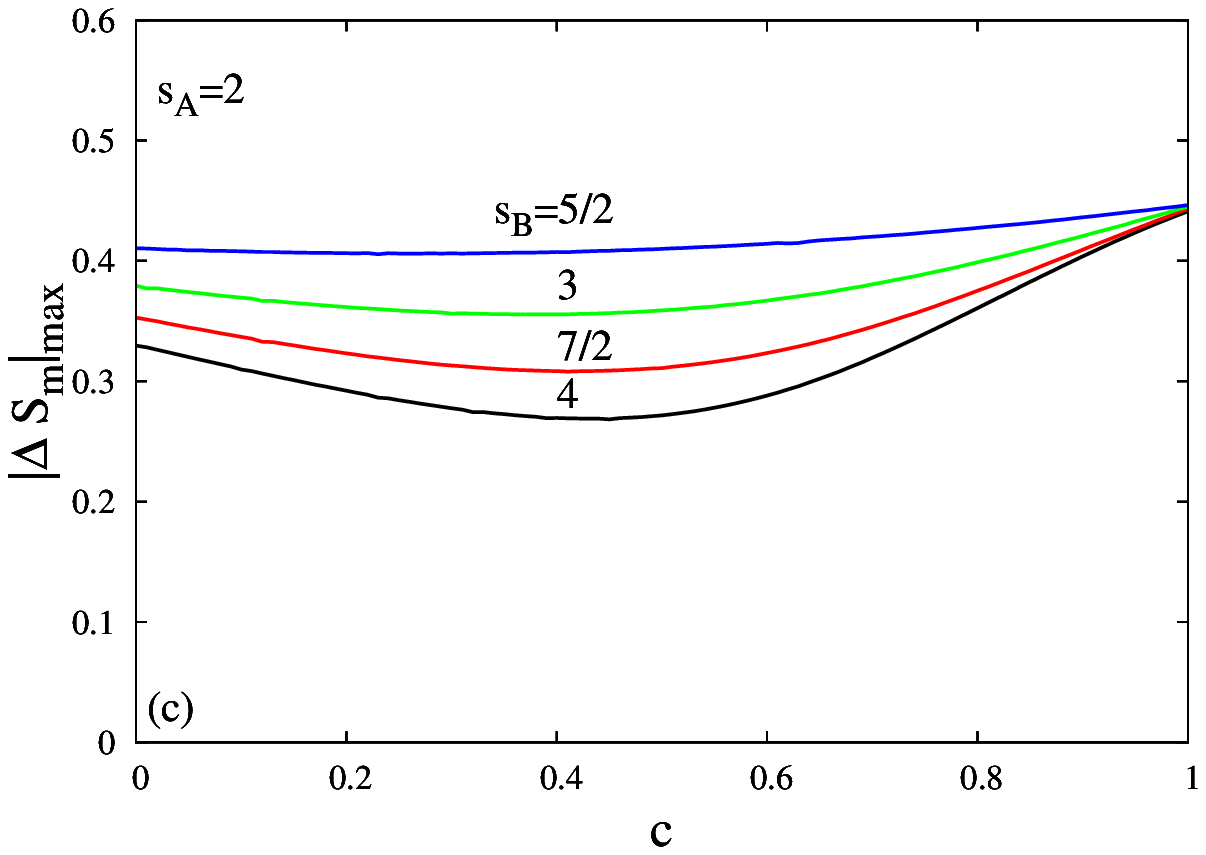, width=6.0cm}
\epsfig{file=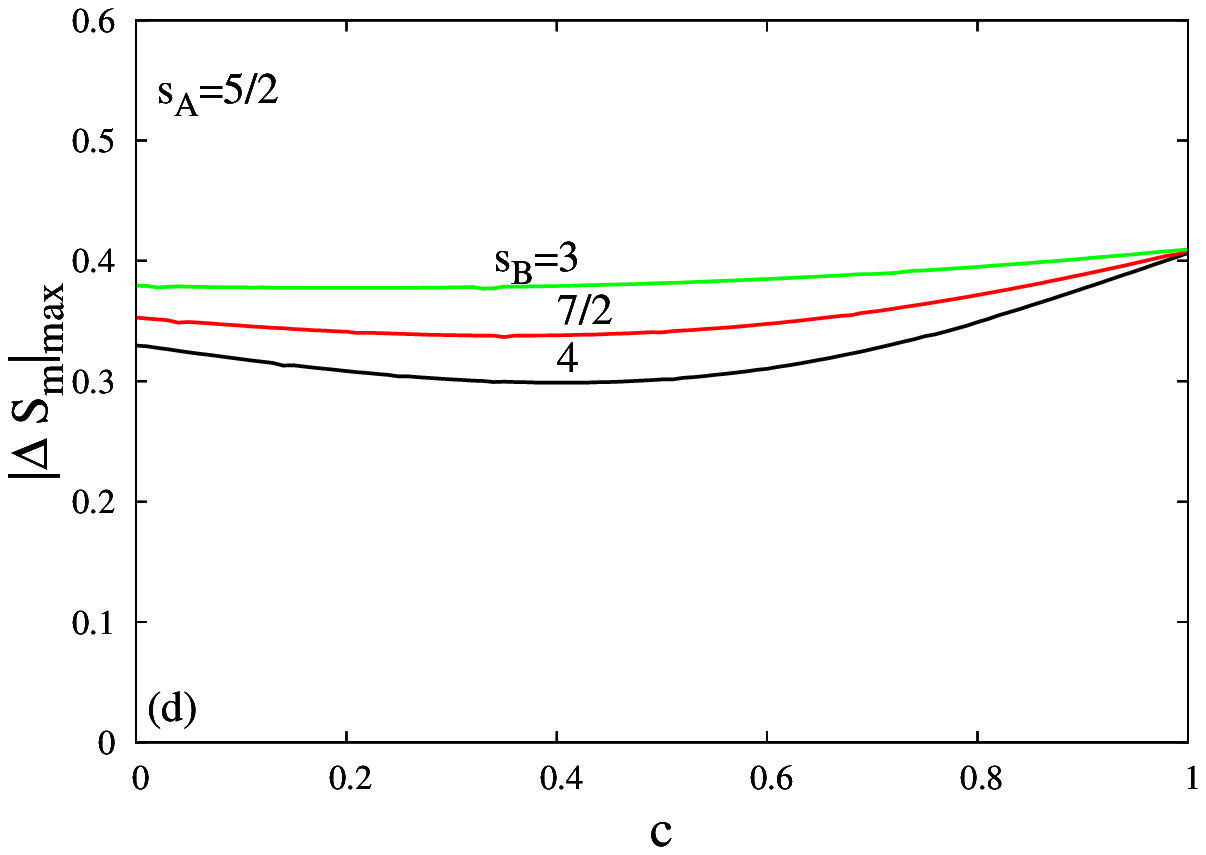, width=6.0cm}
\end{center}
\caption{The variation of the maximum value of the IMEC with the concentration for binary alloy 
$s_A-s_B (S_B>S_A)$ for 
(a) $S_A=1$, (b)$S_A=3/2$, (c)$S_A=2$ and (d)$S_A=5/2$. The values of $S_B$ are given above of the each curve.}
\label{sek2}
\end{figure}

In Fig. \re{sek2} we depict the variation of the  maximum value of the IMEC with the concentration for selected
spin values of $S_A=1, 3/2, 2, 5/2$ and $s_B= 3/2, 2, 5/2, 3, 7/2,4$. As seen in Fig. \ref{sek2}, the behavior of 
the maximum value of the IMEC
by concentration is the same for all the binary alloys: rising concentration first results in decreasing behavior, 
then increasing  behavior takes place while the concentration increases. The final value of the maximum value of the 
IMEC (i.e. at $c=1.0$)
is higher than the initial value (at $c=0.0$) as expected \cite{ref19}. 

\begin{figure}[h]\begin{center}
\epsfig{file=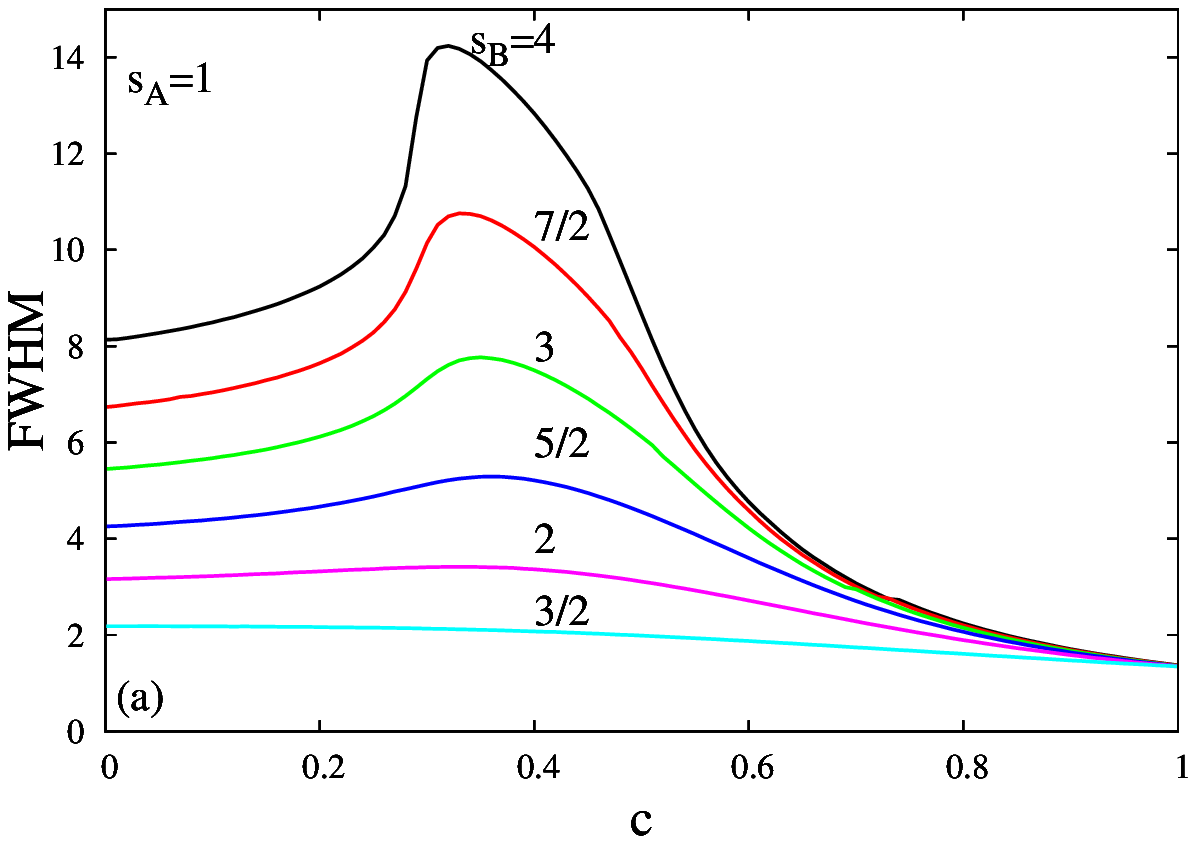, width=6.0cm}
\epsfig{file=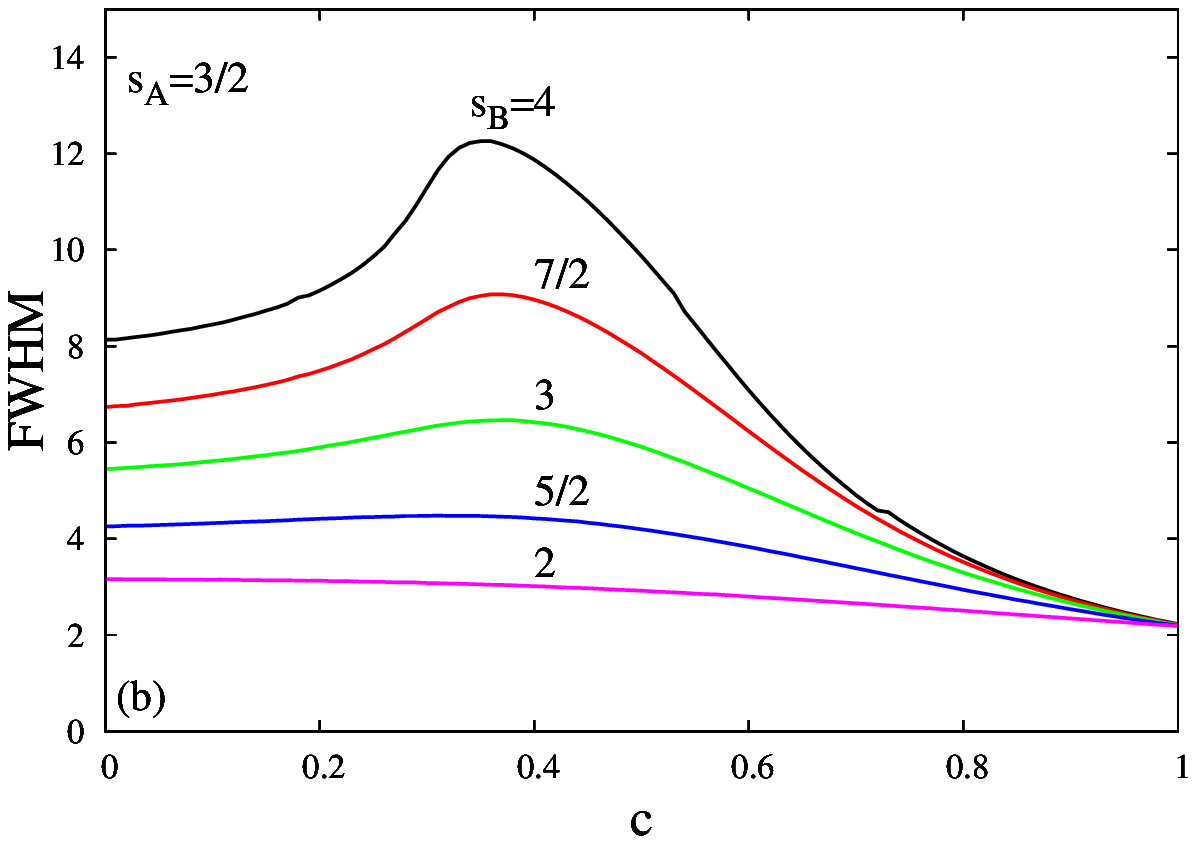, width=6.0cm}
\epsfig{file=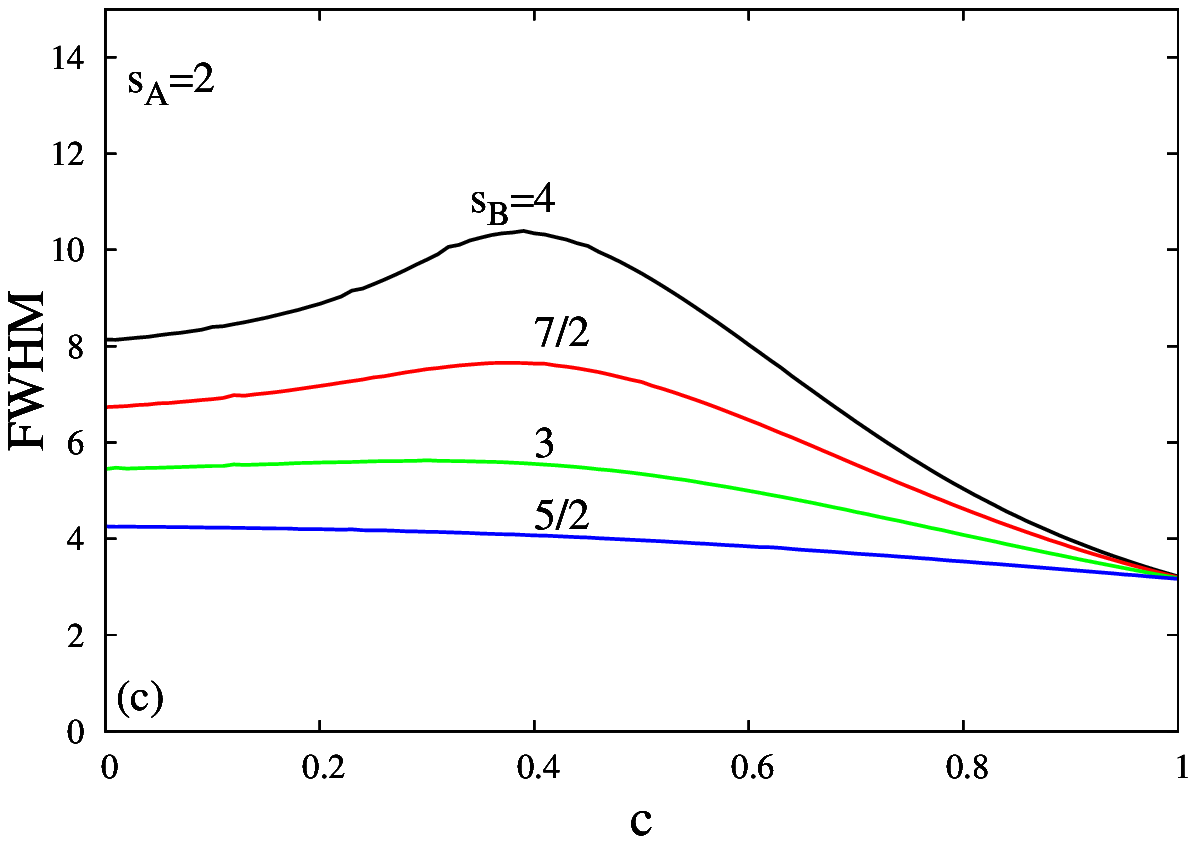, width=6.0cm}
\epsfig{file=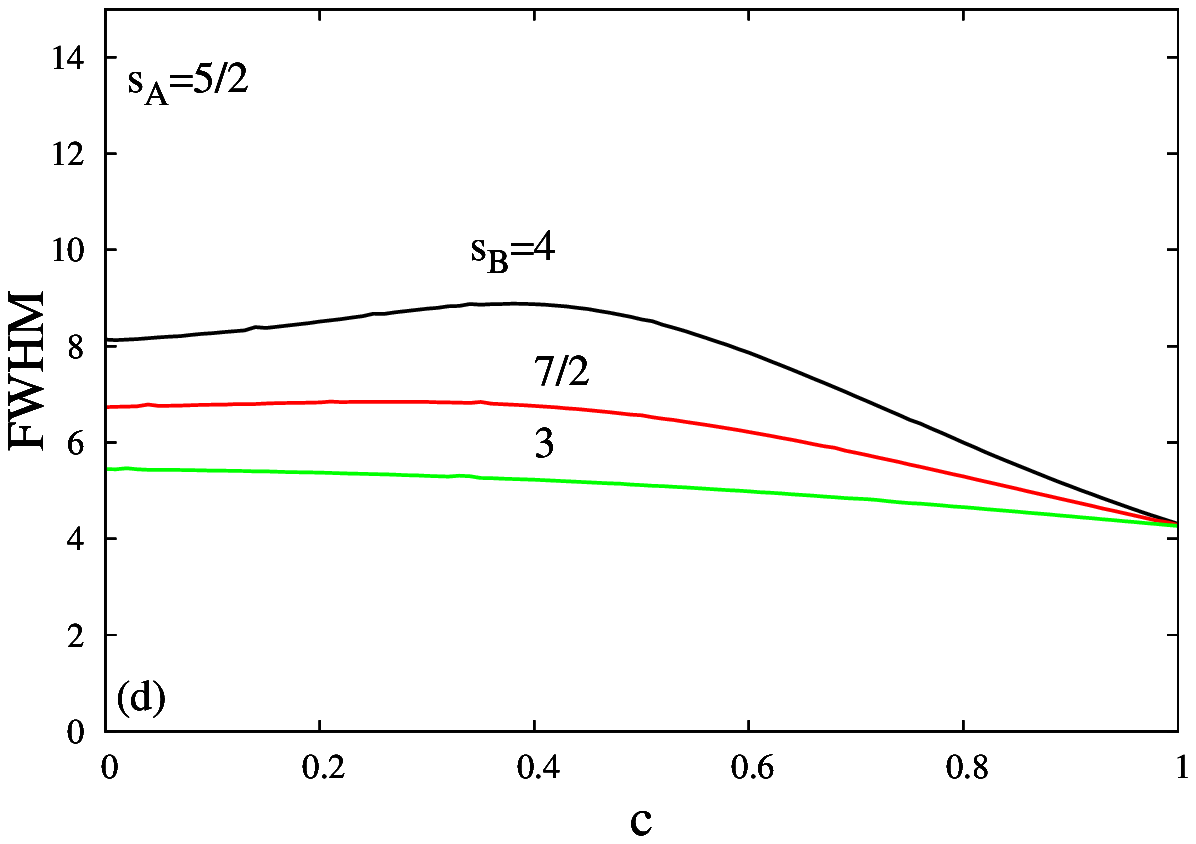, width=6.0cm}
\end{center}
\caption{The variation of the FWHM with the concentration for binary alloy 
$s_A-s_B (S_B>S_A)$ for 
(a) $S_A=1$, (b)$S_A=3/2$, (c)$S_A=2$ and (d)$S_A=5/2$. The values of $S_B$ are given above of the each curve.} 
\label{sek3}
\end{figure}

For investigation of the same behavior 
for FWHM we depict the variation of the FWHM with the concentration in Fig. \ref{sek3}, for 
the same binary alloys shown in Fig \ref{sek2}. As seen in Fig. \ref{sek3},  as expected again, the final value 
of the FWHM is lower than the initial value, when the concentration increases, for all spin pairs. But 
the behavior at the intermediate concentrations is again interesting. While the value of the concentration increases,
first
increment behavior occurs and then decreasing behavior  takes place. This behavior is more evident when 
the binay alloy is constructed from quite different spin values (for example $(S_A,S_B)=(1,4)$).

\begin{figure}[h]\begin{center}
\epsfig{file=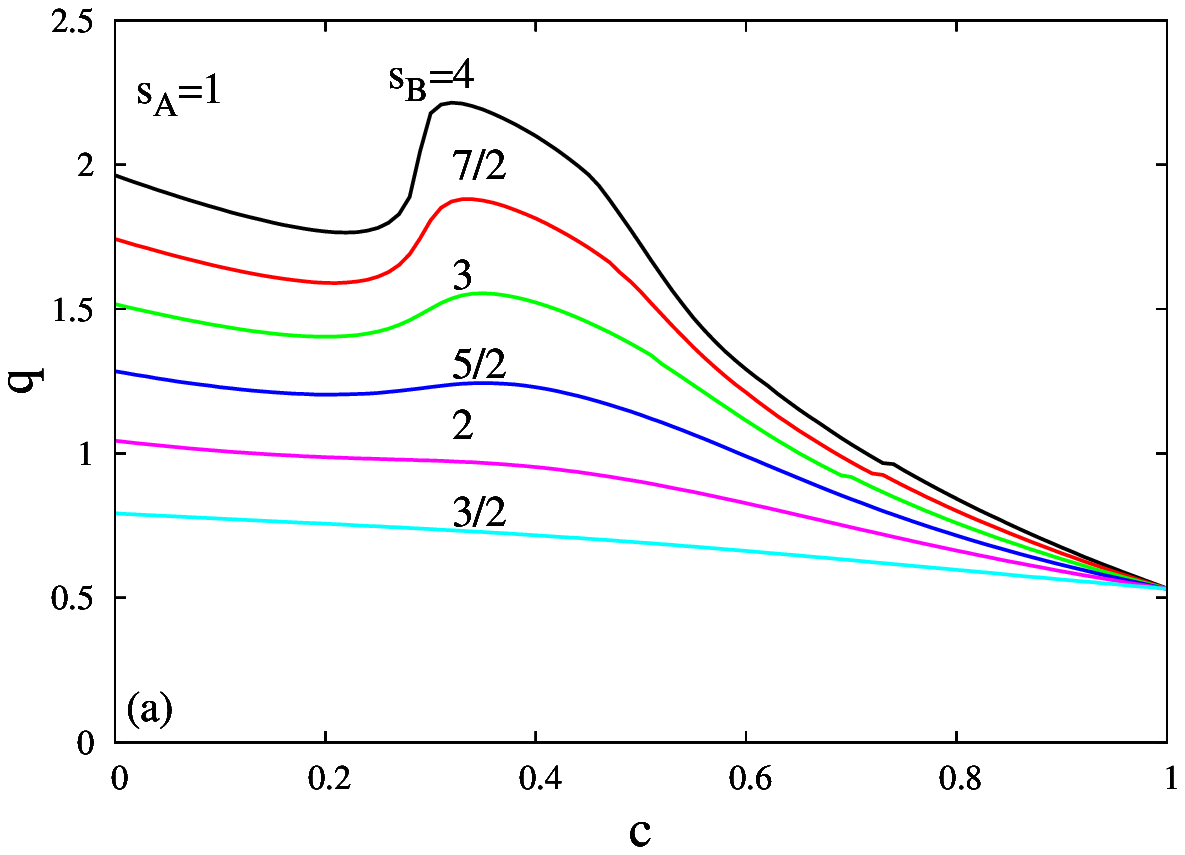, width=6.0cm}
\epsfig{file=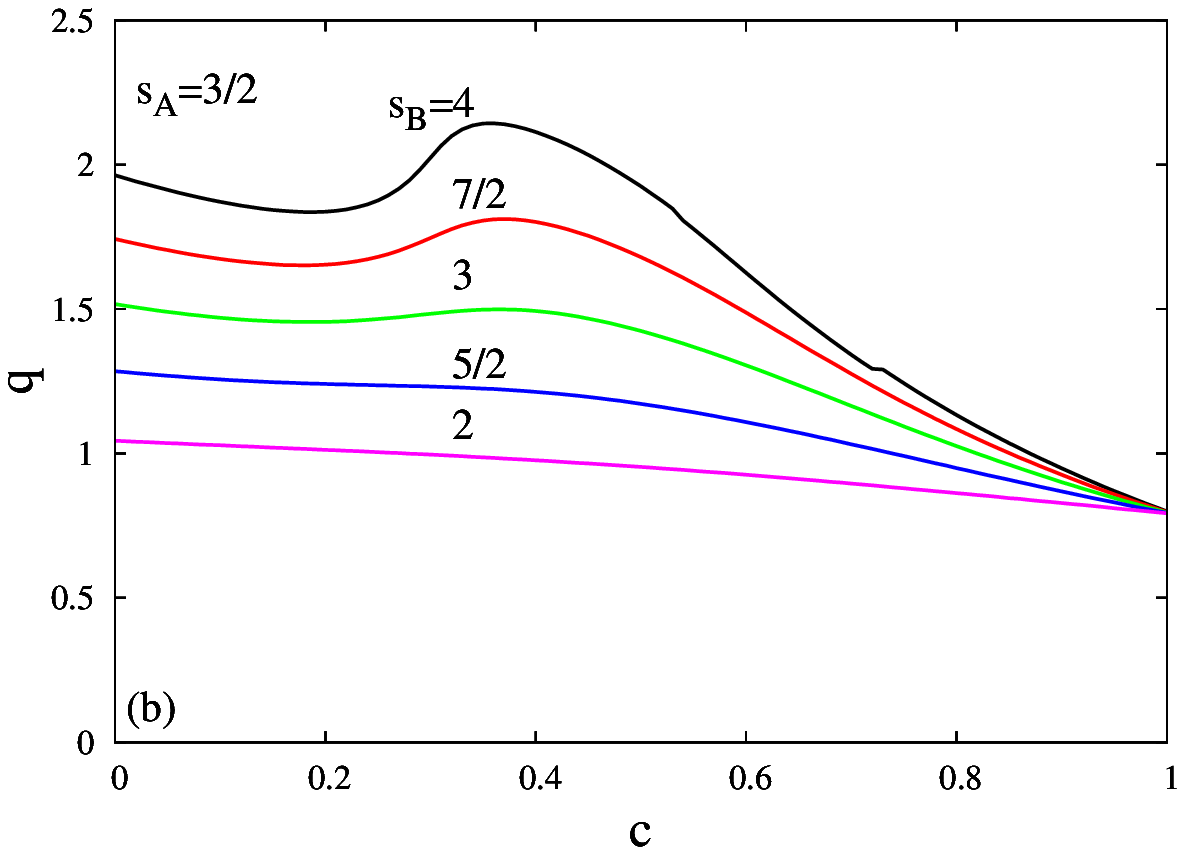, width=6.0cm}
\epsfig{file=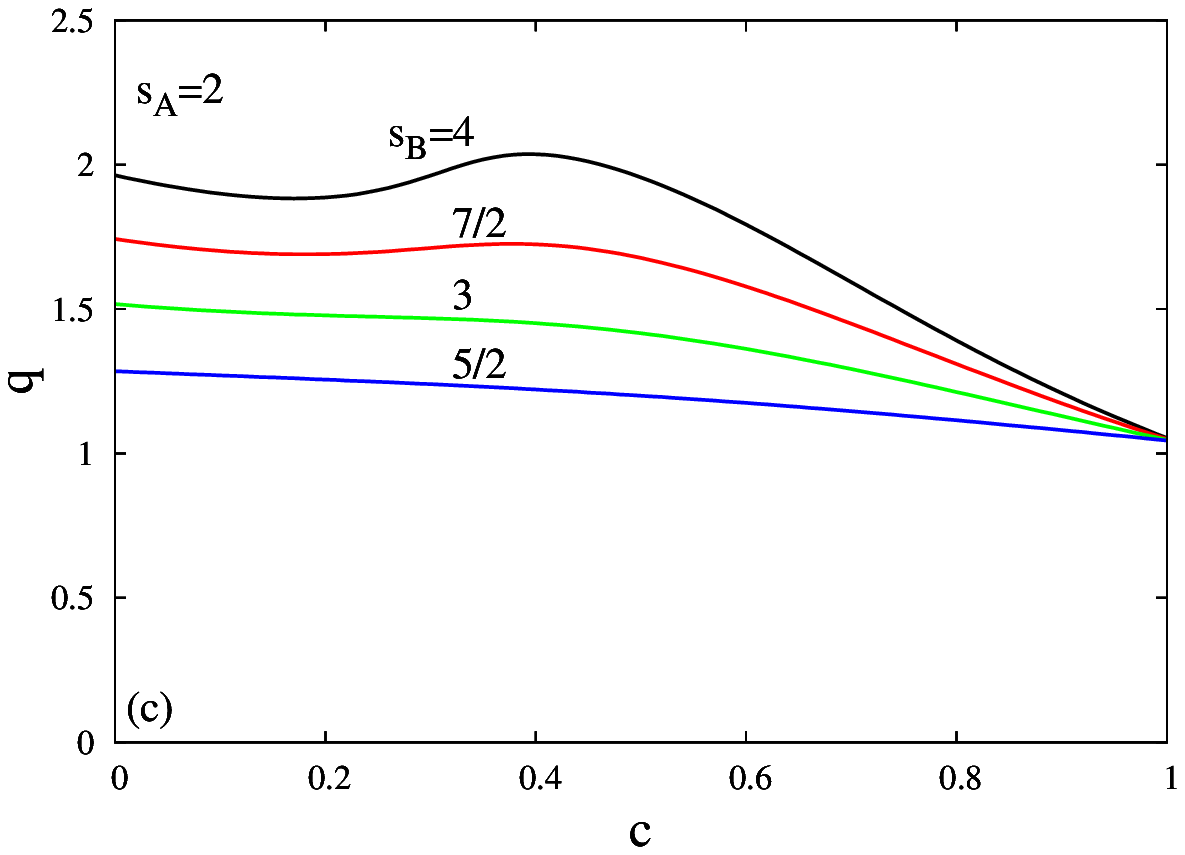, width=6.0cm}
\epsfig{file=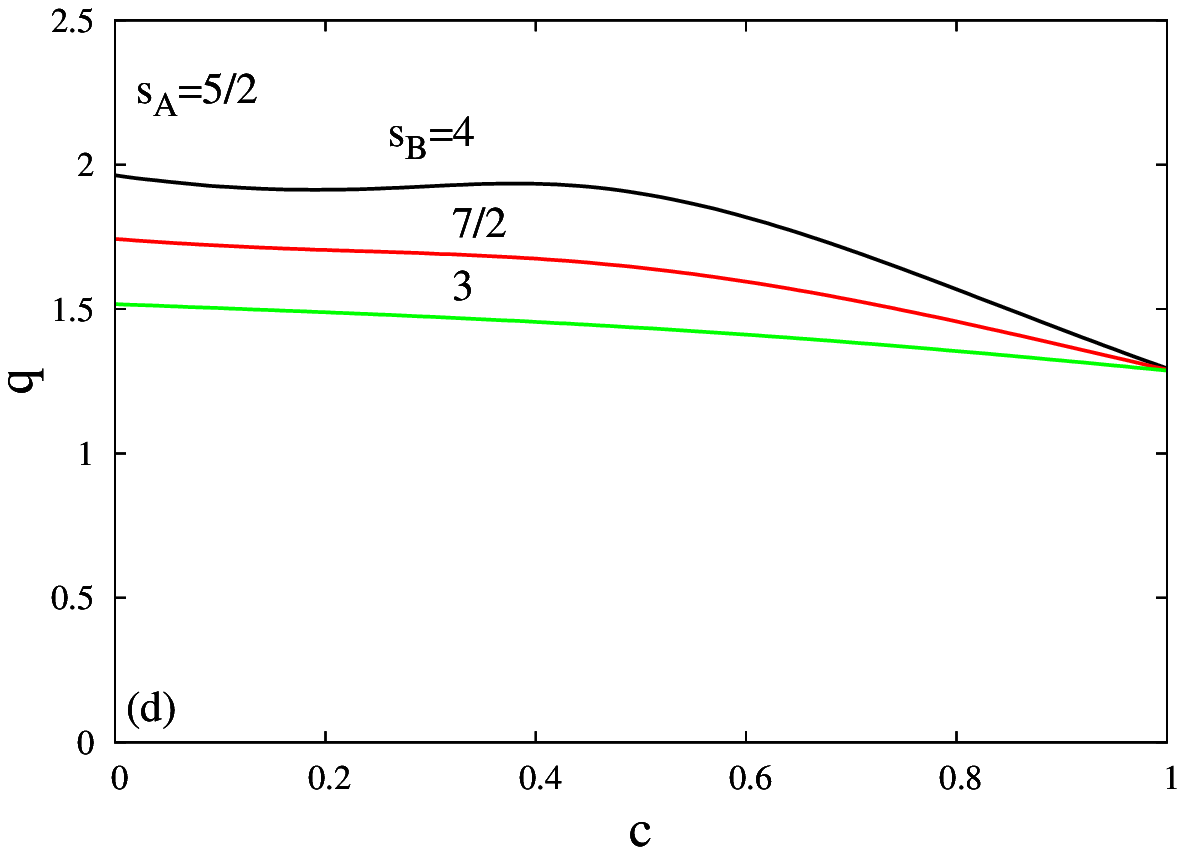, width=6.0cm}
\end{center}
\caption{The variation of the half width of the RC with the concentration for binary alloy 
$s_A-s_B (S_B>S_A)$ for 
(a) $S_A=1$, (b)$S_A=3/2$, (c)$S_A=2$ and (d)$S_A=5/2$. The values of $S_B$ are given above of the each curve.} 
\label{sek4}
\end{figure}

The RC is defined in Eq. \re{denk17} as the area of the portion of the IMEC curve 
between temperature range  of FWHM. Then the behavior of the RC  with the concentration is determined 
by behaviors of the maximum value of the IMEC and FWHM with concentration. We can see
the behavior of $q$ with the concentration in Fig. \ref{sek4}. When we compare the behaviors depicted in
Fig. \ref{sek4} 
by behaviors given in Fig. \ref{sek3} we see that the same characteristics occur. Rising concentration first 
rises $q$, after then decreasing behavior occurs. Again $q$ value of $c=1$ is lower than the 
$q$ value of $c=0$ is compatible with the result obtained before \cite{ref19}.

In order to further elaborate on the  investigation, we give the concentration values that makes the RC 
maximum in Table \re{tbl_1}. As seen in Table \re{tbl_1}, for some combinations, maximum RC occurs at the 
$c=0.0$ i.e. pure system with higher spin. But in contrast to this, for some spin 
combinations the maximum value of the RC occurs at the intermediate concentration values. These binary alloys 
are constructed by using spin values that are substantially different from each other, for instance $(S_A,S_B)=(1,3)$ or
$(S_A,S_B)=(2,4)$. This means that, the value of RC can be tuned by adjusting the concentration value.  

\begin{table}\caption{The concentration values that make the cooling capacity maximum. }\label{tbl_1}
 \begin{tabular}{c|c|c|c|c|c|cc}
 $s_A / s_B$  &3/2&2&5/2&3&7/2&4\\
  \hline
 1&0.00&
0.00&
0.00&
0.34&
0.32&
0.32&
\\
  \hline
 3/2&
&
0.00&
0.00&
0.00&
0.36&
0.34&
\\
  \hline
 2&
 &
 &
0.00&
0.00&
0.00&
0.38&
\\
  \hline
 5/2&

 &
 &
 &
0.00&
0.00&
0.00&
\\
\end{tabular}
\end{table}

\section{Conclusion}\label{conclusion}

The MCE properties of the Ising  binary alloys constituted from arbitrary spin values, 
have been determined by using effective field theory. For determining the efficiency of the MCE  
in binary alloy, IMEC, RC and FWHM quantities have been calculated  for various values of spin and concentration. 

In our earlier work we have determined the relation between the MCE properties and 
spin value for general spin-S Ising system \cite{ref19}. Our conclusions were,  when the spin value increases 
the maximum value of the IMEC decreases and  the FWHM and RC
increase. If we think only in terms of the parameter RC, we can conclude that higher spin materials 
are desirable in order to get more cooling power in MCE. 

In this work we performed same analysis on Ising binary alloys. Two limits of the concentration values 
$c=0$ and $c=1$ gives the pure spin-S Ising model, i.e. these limits 
give the results obtained in our earlier work\cite{ref19}. Surprisingly, we obtained no monotonic behavior 
of the properties of the MCE when the concentration changes. Perhaps most importantly, gretaer RC can be obtained 
for intermediate concentration values, in   comparison with the limiting concentration values. This result allows the 
tunability of the MCE performance by changing concentration.  

We hope that the results  obtained in this work may be beneficial form both 
theoretical and experimental point of view.


\newpage

 \begin{thebibliography}{00}


 
\bibitem{ref1} A.M. Tishin, Y.I. Spichkin, The Magnetocaloric Effect and Its Applications, Institute of Physics, 2003.
\bibitem{ref2} E. Warburg, Ann. Phys. 13 (1881) 141.
\bibitem{ref3} P. Weiss, A. Piccard, J. Phys. (Paris) 7 (1917) 103.
\bibitem{ref4} P. Debye, Ann. Phys. 81 (1926) 1154.


\bibitem{ref5} V. Franco, J.S. Blázquez, B. Ingale, A. Conde, Annu. Rev. Mater. Res. 42 (2012) 305.

\bibitem{ref6} K.A. Gschneidner Jr., V.K. Pecharsky, Annu. Rev. Mater. Sci. 30 (2000) 387.


\bibitem{ref7} J. Lyubina, J. Phys. D, Appl. Phys. 50 (2017) 053002.
\bibitem{ref8} V. Franco, J.S. Blázquez, J.J. Ipus, J.Y. Law, L.M. Moreno-Ramírez, A. Conde, Prog.
Mater. Sci. 93 (2018) 112.

\bibitem{ref9} K. Szalowski, T. Balcerzak, A. Bobák, J. Magn. Magn. Mater. 323 (2011) 2095.
\bibitem{ref10} K. Szalowski, T. Balcerzak, J. Phys. Condens. Matter 26 (2014) 386003.
\bibitem{ref11} M. Topilko, T. Krokhmalskii, O. Derzhko, V. Ohanyan, Eur. Phys. J. B 85 (2012) 278.


 \bibitem{ref12} Y. Ma, A. Du, J. Magn. Magn. Mater. 321 (2009) L65.
 \bibitem{ref13} E.P. Nóbrega, N.A. de Oliveira, P.J. von Ranke, A. Troper, Phys. Rev. B 72 (2005) 134426.

 \bibitem{ref14} R. Arai, R. Tamura, H. Fukuda, J. Li, A.T. Saito, S. Kaji, H. Nakagome, T. Numazawa,
 IOP Conf. Ser., Mater. Sci. Eng. 101 (2015) 012118.



\bibitem{ref15} F. C. S\'{a}Barreto, I. P. Fittipaldi, B. Zeks, Ferroelectrics 
39 (1981) 1103.

\bibitem{ref16} J. Strecka, M. Jascur, acta physica slovaca  65 (2015) 235.

\bibitem{ref17} R. Honmura, T. Kaneyoshi, J. Phys. C: Solid State Phys. 12 
(1979) 3979. 


\bibitem{ref18} T. Kaneyoshi, J. Tucker, M. Jascur, Physica A 176, 495 (1992).



\bibitem{ref19} \"U. Ak\i nc\i \, Y. Y\"uksel, E. Vatansever Physics Letters A 382, 3238 (2018).


 \end{thebibliography}
\end{document}